\documentclass[aps,prl,
preprint,
groupedaddress,amsmath,amssymb]{revtex4-2}
\usepackage{graphicx}% Include figure files
\newcommand{\snn}{\sqrt{s_\text{NN}}}

\begin{document}

\preprint{}

\title{Three-Nucleon Correlations in Light Nuclei Yields Ratios from AMPT Model for QCD Critical Point Investigation}


\author{Ning Yu}
\email{ning.yuchina@gmail.com}
\author{Zuman Zhang}
\author{Hongge Xu}
\author{Zhong Zhu}
\affiliation{School of Physics and Mechanical Electrical and Engineering, Hubei University of Education, Wuhan 430205, China}
\affiliation{Institute of Theoretical Physics, Hubei University of Education, Wuhan 430205, China}
\author{Zhong Zhu}
\affiliation{Institute of Quantum Matter, South China Normal University, Guangzhou, Guangdong 510631}
\affiliation{School of Physics and Mechanical Electrical and Engineering, Hubei University of Education, Wuhan 430205, China}

\date{\today}

\begin{abstract}
This research use the AMPT model in Au+Au collisions to study the influence of the three nucleons correlation $C_{n^2p}$ on the light nuclei yield ratios. It is found that neglecting $C_{n^2p}$ leads to an overestimated relative neutron density fluctuation extraction. Including $C_{n^2p}$ will enhances the agreement with experimental results with higher yield ratios, yet it does not change the energy dependence of the yield ratio. Since there is no first-order phase transition or critical physics in the AMPT model, our work fails to reproduce the experimental energy-dependent peak around $\snn = $20-30 GeV. Our work might offer a baseline for investigating critical physics phenomena using the light nuclei production as a probe. 
\begin{description}\item[PACS numbers]
\verb+25.75.Nq, 24.10.Lx, 24.10.Pa+
\end{description}
\end{abstract}

%\keywords{}

\maketitle
%\section{Introduction}
Quantum Chromodynamics (QCD), the bedrock theory of strong interactions governing quarks and gluons, drives inquiries into the QCD phase diagram~\cite{RN168}, mapping out the behavior of QCD matter under extreme conditions. One of the pivotal objectives of the Beam Energy Scan (BES) program at the Relativistic Heavy-Ion Collider (RHIC) is the search for the elusive QCD critical point~\cite{RN177,RN178,RN179,RN180,RN181,RN182}. This is also a key physics motivation for future accelerators, such as the Facility for Anti-Proton and Ion Research (FAIR) in Darmstadt and the Nuclotron-based Ion Collider fAcility (NICA) in Dubna.

Close to the QCD critical point, expectations are that fluctuations in conserved quantities, notably baryon number ($B$), charge ($Q$), and strangeness ($S$)~\cite{RN302}. The production of light nuclei is predicted to be sensitive to the baryon density fluctuations, under the premise that these nuclei are formed by the coalescence of nucleons~\cite{RN9,RN118}. The light nuclei yield ratio, expressed as $N_pN_t/N_d^2$, which encompasses the production of proton($p$), deuteron($d$), and triton($t$), can be posited to be delineated by the relative neutron density fluctuation ($\Delta\rho_n$) and the correlation between neutron and proton densities at the kinetic freeze-out. Notably, the STAR collaboration has reported a non-monotonic energy dependence of the yield ratio, peaking around 20-30 GeV, in the most central Au+Au collisions~\cite{RN172,RN218}. If correlations between neutron and proton densities are disregarded, the light nuclei yield ratio is indicative of a direct proportionality with the relative neutron density fluctuations. Hence, the experimental observation implies the existence of a large relative neutron density fluctuation at this energy range.

In our previous work~\cite{RN217}, utilizing the AMPT model, we investigated the impact of the two-body neutron-proton density correlation, $C_{np}$, on the yield ratio of light nuclei, arriving at the conclusion that the correlation  $C_{np}$ has little effect on the light nuclei yield ratio at central or mid-central Au+Au collisions. In other words, experimentalist can extract the relative neutron density fluctuation directly from light nuclei yield ratio. While at peripheral collision, the effect of $C_{np}$ on the light nuclei yield ratio becomes larger, and the related effect must be taken into account when extracting the density fluctuation. However, a critical aspect overlooked in that study concerned the three-nucleon correlation involving two neutrons and one proton, $C_{n^2p}$, which has a direct influence on the triton yields and, consequently, the overall light nuclei yield ratio. Given that tritons are products of coalescence processes involving multiple nucleons, the inclusion of $C_{n^2p}$ is pivotal for a comprehensive understanding of light nuclei formation dynamics and the accurate extraction of the relative neutron density fluctuations from experimental data.

Therefore, in this paper, we aim to build upon our previous findings by delving into the three-nucleon correlation $C_{n^2p}$ on the light nuclei yield ratio. 
Through this enhanced analysis, we expect to show the importance of $C_{n^2p}$ on the extraction of relative neutron density fluctuation from the light nuclei yield ratio, thereby offering insights for the quest of identifying the QCD critical point. This paper is structured as follows: we commence with a review of the AMPT model. We then show the definition of the three-nucleon correlation and its connection to the light nuclei yield ratio. Following this, we present our results on the three-nucleon correlation's dependence on the rapidity coverage, collision centrality and energy. Finally, we discuss the implications of the $C_{n^2p}$ on the light nuclei yield ratio and its observed energy-dependent behavior in experiments. 

The AMPT, a multi-phase transport model is a hybrid model consisting of four components, the initial conditions, partonic interactions, conversion from partonic to hadronic matter, and hadronic interactions~\cite{RN303}. The default version of the AMPT involves only mini-jet partons in the parton cascade and uses the Lund string fragmentation for parton hadronization~\cite{RN305}. On the other hand, the string melting version of the AMPT model, where all the excited strings are converted to partons and a quark coalescence model is used to describe the parton hadronization. Typically, the default version gives a reasonable description of $dN/d\eta$, $dN/dy$, and the $p_T$ spectra, while the string melting version describes the magnitude of the elliptic flow but not the $p_T$ spectra. The string melting version, with a modified set of parameters, can well reproduce the $p_T$ spectra and elliptic flows at RHIC top energy~\cite{RN304}. In this paper, all the results are studied by using this set of parameters.

Base on the references~\cite{RN9,RN172} and our preceding work~\cite{RN217}, we commence with a review of the nucleon coalescence model and its consequent estimations for light nuclear yields. Ignoring the binding energy of light nuclei, their abundance can be formulated as follows:
\begin{widetext}
    \begin{equation}
    N_c=g_cA^{3/2}\left(\frac{2\pi}{m_0T_{\rm eff}}\right)^{3(A-1)/2}V\langle\rho_p\rangle^{A_p}\langle\rho_n\rangle^{A_n}\sum_{i=0}^{A_p}\sum_{j=0}^{A_n}C_{A_p}^iC_{A_n}^jC_{n^jp^i}
\end{equation}
%\end{widetext}
Here, $g_c =\dfrac{2S+1}{2^A}$ represents the coalescence factor for for $A=A_n+A_p$ nucleons of spin $1/2$ forming a cluster with total spin $S$. The nucleon mass $m_0$ is considered equal for both protons and neutrons. $V$ denote the system volum, and $T_{\rm eff}$ is the effective temperature at kinetic freeze-out. $\langle\rho_n\rangle$ and $\langle\rho_p\rangle$ are the neutron and proton density. Combinations are represented by $C_{A_p}^i$ and $C_{A_n}^i$. $C_{n^jp^i}$ is the corelation between $j-$neutrons and $i-$protons, defined as:
\begin{equation}
    C_{n^jp^i}=\frac{\langle\delta\rho_p^i\delta\rho_n^j\rangle}{\langle\rho_p\rangle^i\langle\rho_n\rangle^j}
\end{equation}

The relative neutron density fluctuation $\Delta \rho_n = \sigma_n^2/\langle\rho_n\rangle^2$ is equivalent to $C_{n^2p^0}$. The two-body neutron-proton density correlation, $C_{np}$, is given by:
\begin{equation}
    C_{np}=\frac{\langle\delta\rho_p\delta\rho_n\rangle}{\langle\rho_p\rangle\langle\rho_n\rangle}=\frac{\left\langle\rho_p\rho_n\right\rangle}{\langle\rho_p\rangle\langle\rho_n\rangle}-1
\end{equation}

The three-nucleon correlation, $C_{n^2p}$, can be expressed as
\begin{equation}
    C_{n^2p}=\frac{\langle\delta\rho_p\delta\rho_n^2\rangle}{\langle\rho_p\rangle\langle\rho_n^2\rangle}=\frac{\left\langle\rho_p\rho_n^2\right\rangle}{\langle\rho_p\rangle\langle\rho_n\rangle}-(1+2C_{np})
\end{equation}

Employing these formulations, the yields of deuteron and triton are specified as:
%\begin{widetext}
    \begin{eqnarray}
    N_d&=&\frac{3}{2^{1/2}}\left(\frac{2\pi}{m_0T_{\rm eff}}\right)^{3/2}V\langle\rho_p\rangle\langle\rho_n\rangle(1+C_{np})\\
    N_t&=&\frac{3^{3/2}}{4}\left(\frac{2\pi}{m_0T_{\rm eff}}\right)^{3}V\langle\rho_p\rangle\langle\rho_n\rangle^{2}(1+\Delta\rho_n+2C_{np}+C_{n^2p})
\end{eqnarray}
%\end{widetext}

Subsequently, the light nuclei yields ratio is compactly represented by:
\begin{equation}
    R = \frac{1}{2\sqrt{3}}\frac{1+\Delta\rho_n+2C_{np}+C_{n^2p}}{(1+C_{np})^2}
    \label{lnr}
\end{equation}
It can be observed that the three-nucleon correlation $C_{n^2p}$ exerts a significant influence on the light nuclei yield ratios, effectively enhancing them. Assuming the three-nucleon correlation $C_{n^2p}$ is zero, Eq.~\ref{lnr} simplifies to:
\begin{equation}
    R = \frac{1}{2\sqrt{3}}\frac{1+\Delta\rho_n+2C_{np}}{(1+C_{np})^2}
    \label{lnr1}
\end{equation}
Taking the analysis a step further, if we also neglect the two-nucleon correlation $C_{np}$ in our calculations, the light nuclei yield ratio becomes even more simplified, expressing as:
\begin{equation}
    R = \frac{1+\Delta\rho_n}{2\sqrt{3}}
    \label{lnr2}
\end{equation}
In this highly simplified scenario, the yield ratio is directly proportional to the relative neutron density fluctuation $\Delta\rho_n$, which forms the experimental basis for extracting the $\Delta\rho_n$ from the yield ratios of light nuclei.
\end{widetext}
Following the procedure in our preceding work, the event-by-event multiplicity and fluctuation of proton $\langle N_p\rangle$, $S_p$, neutron $\langle N_n\rangle$, $S_n$ and their mixed moments $\langle N_pN_n\rangle,\langle N_pN_n^2\rangle$ can be extracted from AMPT. In calculating $\Delta \rho_p, \Delta \rho_n$, the system volume effects are canceled out. In AMPT model, nucleon production is analyzed across varying rapidity intervals and collision centralities. The definition of centrality is determined by the per-event charged particle multiplicity $N_{\rm ch}$ for pseudorapidity range $\eta\leqslant$ 0.5. 

\begin{figure}[tb]
    \includegraphics[width=0.8\textwidth]{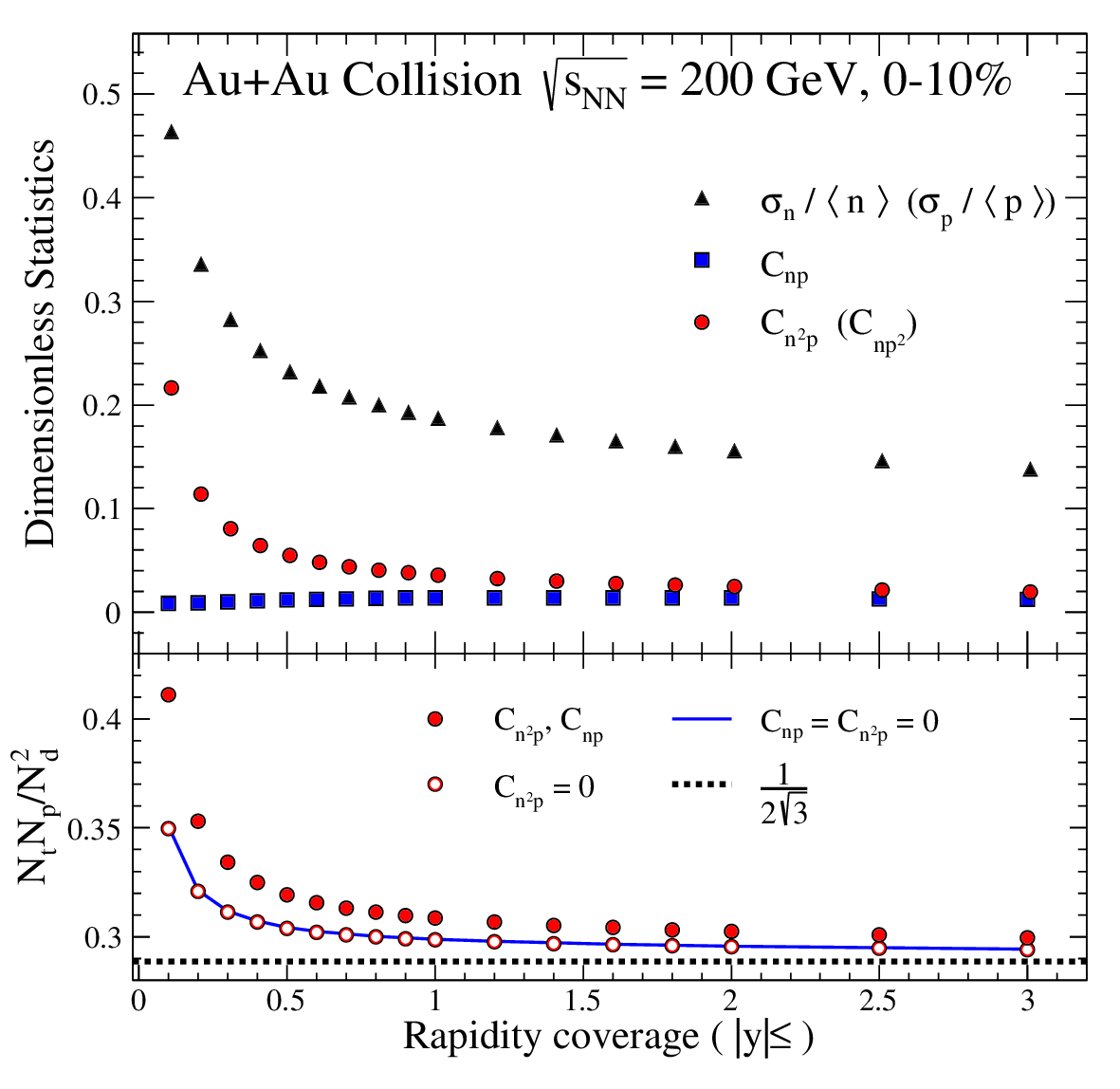}
    \caption{\label{fig1} Top panel: Dimensionless statistics $\sigma_n/\langle n \rangle$, $\sigma_p/\langle p \rangle$, $C_{np}$, $C_{n^2p}$, and $C_{np^2}$ for $0-10\%$ Au+Au collisions at $\snn=200$ GeV. Bottom panel: The light nuclei yield ratio $N_tN_p/N_d^2$ calculated from top panel are shown as solid circles by Eq.~(\ref{lnr}), opened circles by Eq.~(\ref{lnr1}) and solid line by Eq.~(\ref{lnr2}).}
\end{figure}

Fig.~\ref{fig1} illustrates the rapidity coverage dependence of dimensionless statistics $\sigma_n/\langle n \rangle$, $\sigma_p/\langle p \rangle$, $C_{np}$, $C_{n^2p}$, and $C_{np^2}$ for $0-10\%$ Au+Au collisions at $\snn=200$ GeV. $\sigma_p/\langle\rho_p\rangle$ and $\sigma_n/\langle\rho_n\rangle$, which can be regarded as relative nucleon density fluctuations, decrease with increasing rapidity coverage. It can be found that the relative density fluctuation for neutrons $\sigma_n/\langle n \rangle$ and protons $\sigma_p/\langle p \rangle$ are roughly equivalent and exhibit a decline as the rapidity coverage increasing. In the smaller rapidity coverage region, especially at mid-rapidity, particle pair production dominates. As a result, nucleon density fluctuations are relatively larger at mid-rapidity compared to a wider rapidity coverage. The correlation $C_{np}$ is independent of rapidity coverage and almost vanished for 0-10\% Au+Au collisions at $\snn = $ 200 GeV. A similarity is observed between the correlation $C_{n^2p}$ for two neutrons and one proton and the correlation $C_{np^2}$ for one neutron and two protons. The behavior of $C_{n^2p}$ is similar to that of the relative neutron density fluctuations $\sigma_n/\langle n \rangle$, both decreasing as the rapidity coverage increases.

The lower panel of Fig.~\ref{fig1} presents the computed results of the light nuclei yield ratios derived from Eq.~(\ref{lnr}), ~(\ref{lnr1}), and (\ref{lnr2}), respectively demonstration the comprehensive influence of both both $C_{np}$ and $C_{n^2p}$, the isolates effect of $C_{np}$ without $C_{n^2p}$, and the scenario devoid of any nucleon correlation effects $C_{np}$ and $C_{n^2p}$. We also draw the line of $1/2\sqrt{3}$, which means both relative neutron density fluctuation and nucleon correlations being vanished. It is observed that, due to the near-zero value of $C_{np}$ for central collisions, its impact on the light nuclei yield ratios is insignificant. Moreover, since $C_{n^2p}$ exhibits a similar dependence on the rapidity coverage as the relative neutron density fluctuation $\sigma_n/\langle n \rangle$, including $C_{n^2p}$ leads to an overall enhancement in the calculated of the light nuclei yield ratios. Consequently, if $C_{n^2p}$ is disregarded in the analysis, employing Eq.~(\ref{lnr1}) to extract the relative neutron density fluctuation from the light nuclei yield ratios would yield an overestimated value compared compared to the true physical situation.

\begin{figure}[tb]
    \includegraphics[width=0.8\textwidth]{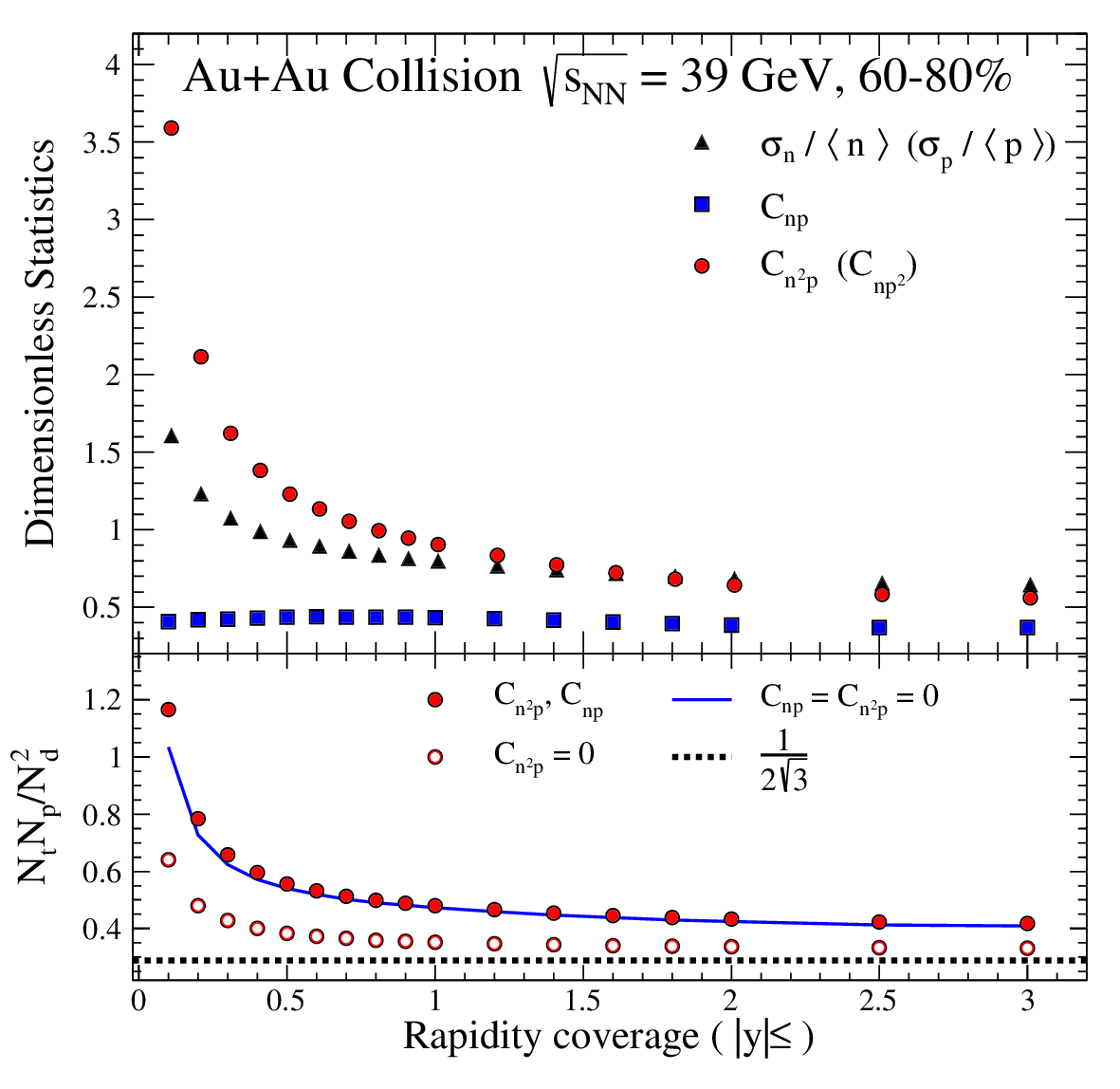}
    \caption{\label{fig2} Top panel: Dimensionless statistics $\sigma_n/\langle n \rangle$, $\sigma_p/\langle p \rangle$, $C_{np}$, $C_{n^2p}$, and $C_{np^2}$ for $60-80\%$ Au+Au collisions at $\snn=39$ GeV. Bottom panel: The light nuclei yield ratio $N_tN_p/N_d^2$ calculated from top panel are shown as solid circles by Eq.~(\ref{lnr}), opened circles by Eq.~(\ref{lnr1}) and solid line by Eq.~(\ref{lnr2}).}
\end{figure}

The top panel of Fig.~\ref{fig2} shows the rapidity coverage dependence of $\sigma_n/\langle n \rangle$, $\sigma_p/\langle p \rangle$, $C_{np}$, $C_{n^2p}$, and $C_{np^2}$ for $60-80\%$ Au+Au collisions at $\snn=39$ GeV. Consistent with the findings from central collisions, $\sigma_p/\langle\rho_p\rangle$, $\sigma_n/\langle\rho_n\rangle$, and $C_{np^2}$ decrease with increasing rapidity coverage. At a given rapidity coverage, these quantities are greater in peripheral collisions compared to those in central collisions. For instance, the converge value of $\sigma_p/\langle\rho_p\rangle$ at larger rapidity coverage is approximately 0.7 for peripheral collisions, whereas it is roughly 0.14 for cental collision at $\snn=39$ GeV. The $C_{np}$ is independent of rapidity coverage with a non-negligible value about 0.45-0.5 in peripheral collisions. Consequently, the influence of both $C_{np}$ and $C_{n^2p}$ on the related light nuclei yield ratio in peripheral collisions is evident in the bottom panel of Fig.~\ref{fig2}. Specifically, the exclusion of $C_{n^2p}$ is shown to yield a diminished light nuclei yield ratio, that is a change from solid circles to opened circles. Conversely, when $C_{np}$ is not considered, the light nuclei yield ratio is observed to increase, reflected by the shift from opened circles to solid line. It can be observed that the impacts of $C_{np}$ and $C_{n^2p}$ on the light nuclei yield ratio are contrasting. When both are taken into account, the deviation in the yield ratio from that obtained by neglecting both depends critically on the magnitude of their respective effects. Notably, the figure illustrates that for $60-80\%$ Au+Au collisions at $\snn=39$ GeV, the influences of these two factors nearly cancel each other out. In other words, in this particular case, considering both $C_{np}$ and $C_{n^2p}$ together yields results similar to those derived when neither is considered. This highlights that, at least in this instance, the net effect of incorporating $C_{np}$ and $C_{n^2p}$ in the analysis does not significantly change the light nuclei yield ratio.

\begin{figure}[tb]
    \includegraphics[width=0.8\textwidth]{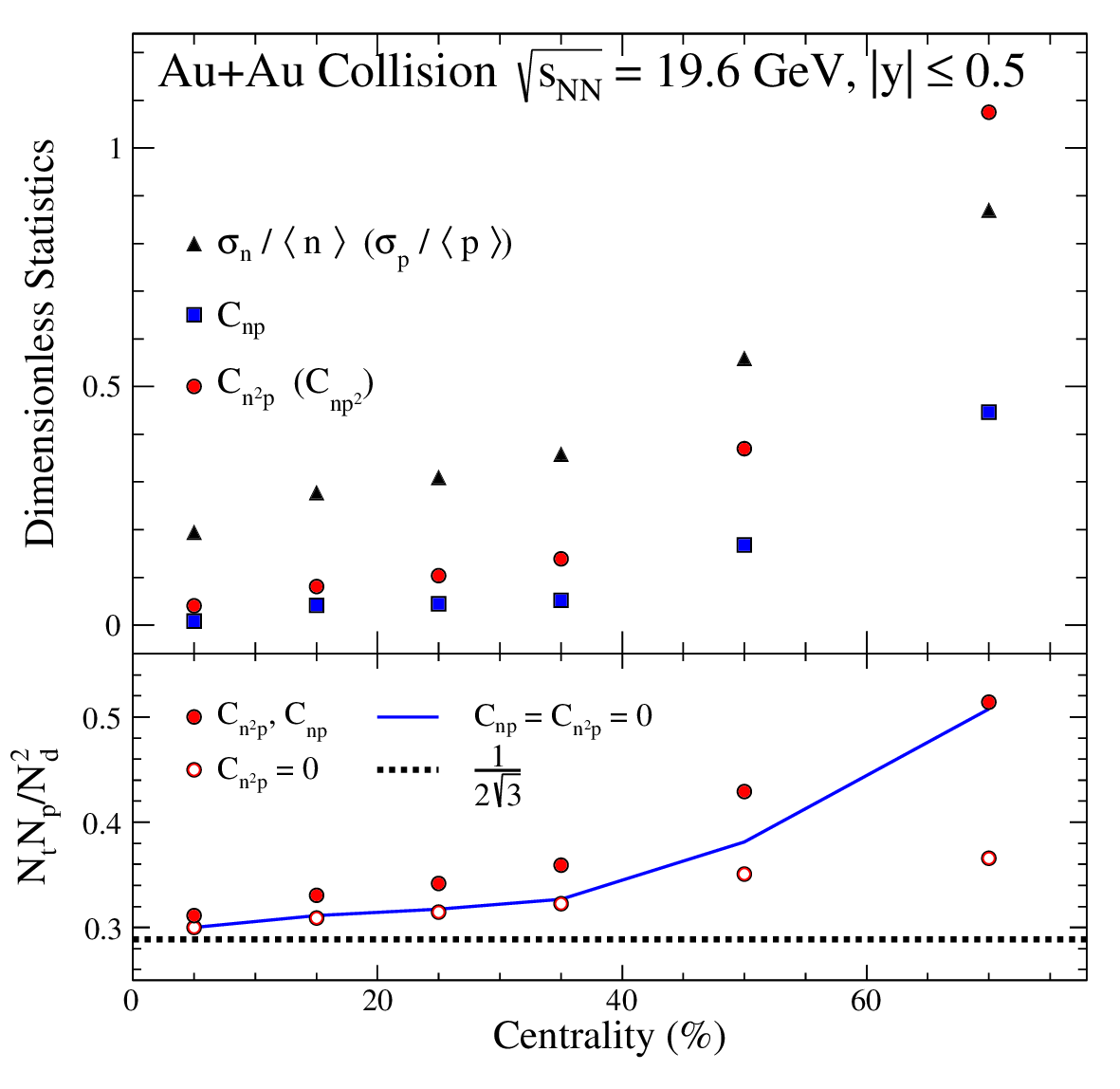}
    \caption{\label{fig3} Top panel: Dimensionless statistics $\sigma_n/\langle n \rangle$, $\sigma_p/\langle p \rangle$, $C_{np}$, $C_{n^2p}$, and $C_{np^2}$ for Au+Au collisions at $\snn=19.6$ GeV with $|y|\leqslant 0.5$. Bottom panel: The light nuclei yield ratio $N_tN_p/N_d^2$ calculated from top panel are shown as solid circles by Eq.~(\ref{lnr}), opened circles by Eq.~(\ref{lnr1}) and solid line by Eq.~(\ref{lnr2}).}
\end{figure}

The top panel of Fig.~\ref{fig3} illustrates the centrality dependence of $\sigma_n/\langle n \rangle$, $\sigma_p/\langle p \rangle$, $C_{np}$, $C_{n^2p}$, and $C_{np^2}$ for Au+Au collisions at $\snn=19.6$ GeV, confined to a rapidity coverage of $|y|\leqslant 0.5$. Notably, these quantities exhibit an increasing trend as the collisions transition from central to peripheral collisions. The corresponding light nuclei yield ratio calculated by Eq.~(\ref{lnr}), Eq.~(\ref{lnr1}), and  Eq.~(\ref{lnr2}) are presented in the bottom of Fig.~(\ref{fig3}). These illustrate the influence of $C_{np}$ and $C_{n^2p}$ on these yields. It is found at central or mid-central collision, the variation between the yield ratios given by Eq.~(\ref{lnr1}) and Eq.~(\ref{lnr2}) is very small, implying that the influence of $C_{np}$ on the yield ratios can effectively be disregarded in these collisions. Conversely, in peripheral collisions, the effect of $C_{np}$ on the yield ratios becomes significant and cannot be ignored. Furthermore, it is observed that $C_{n^2p}$ exerts a positive effect on the yield ratios, thereby causing an increment in the light nuclei yield ratio. In contrast, $C_{np}$ exerts a negative effect on the yield ratios. Consequently, when both $C_{np}$ and $C_{n^2p}$ are taken into account, in central and mid-central collisions, the dominant impact comes from $C_{n^2p}$, leading to an increase in the yield ratio. However, in peripheral collisions, the influences of these two factors may cancel each other. 

\begin{figure}[!htp]
    \includegraphics[width=0.8\textwidth]{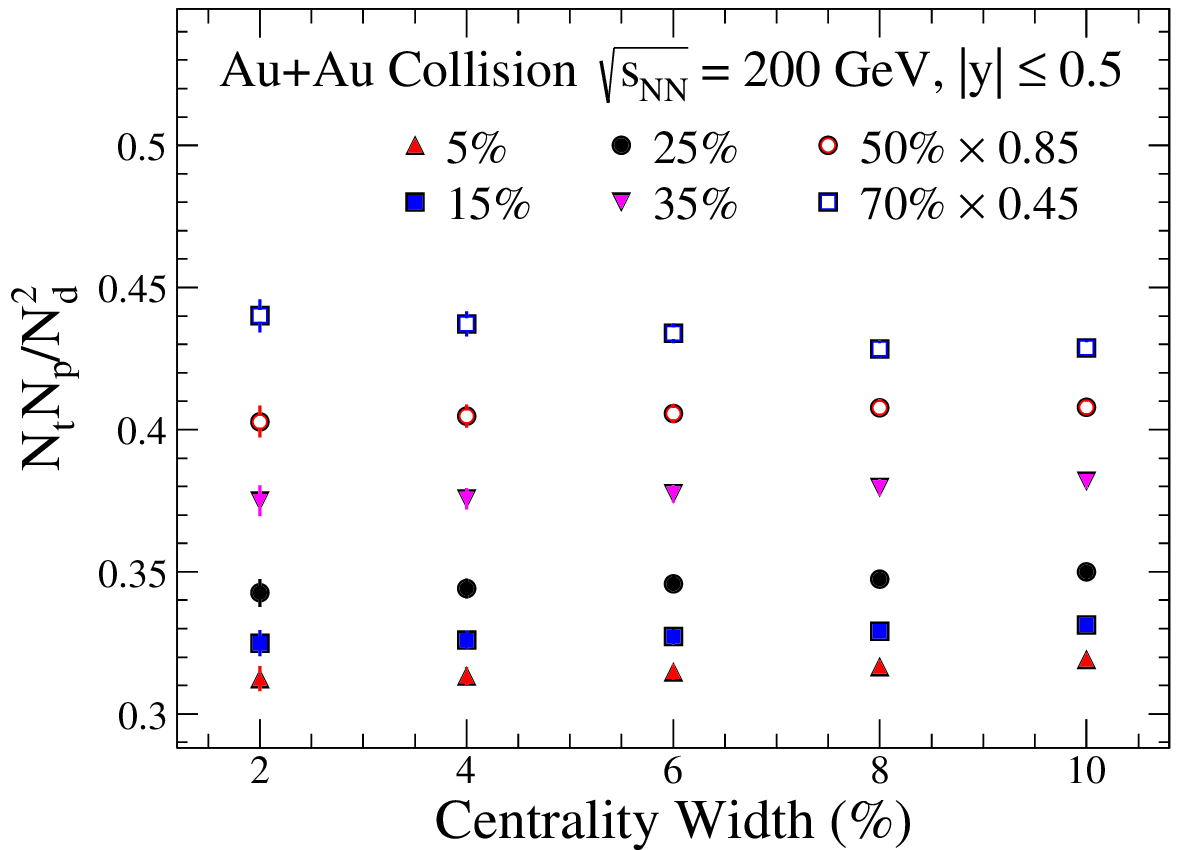}
    \caption{\label{bw} Centrality bin width dependence of the light nuclei yield ratio $N_tN_p/N_d^2$ from AMPT for Au+Au collisions at $\snn=200$ GeV with $|y|\leqslant 0.5$. Different symbols represent results obtained at different centrality centered around specific centrality values (5\%, 15\%, 25\%, 35\%, 50\%, 70\%). Data points for centrality centers of 50\% and 70\% are scaled by factors of 0.85 and 0.45, respectively, for clarity.}
\end{figure}

Centrality bin width correction is important for any event-by-event fluctuation calculation. In Fig.~\ref{bw}, we present the light nuclei yield ratios obtained from different centrality bin widths centered around six centrality. The x-axis represents the centrality bin width. For example, for the red solid triangles, a centrality width of 2\% corresponds to collisions within the centrality of 6\%-7\%, while a centrality width of 5\% corresponds to collisions within the centrality of 0\%-10\%, representing the central collisions reported in this study. Similarly, for the opened circles, a centrality width of 6\% indicates the centrality of 47\%-53\%. To better illustrate the results, the data points corresponding to centrality centers of 50\% and 70\% are multiplied by factors of 0.85 and 0.45, respectively. As shown in the figure, except for the most peripheral collisions, for a given centrality center value, the light nuclei yield ratios increase slightly with increasing centrality width, with the increase being approximately between 2\% to 4\%. This suggests that the influence of centrality bin width on the light nuclei yield ratios is relatively minor.

\begin{figure}[tb]
    \includegraphics[width=0.8\textwidth]{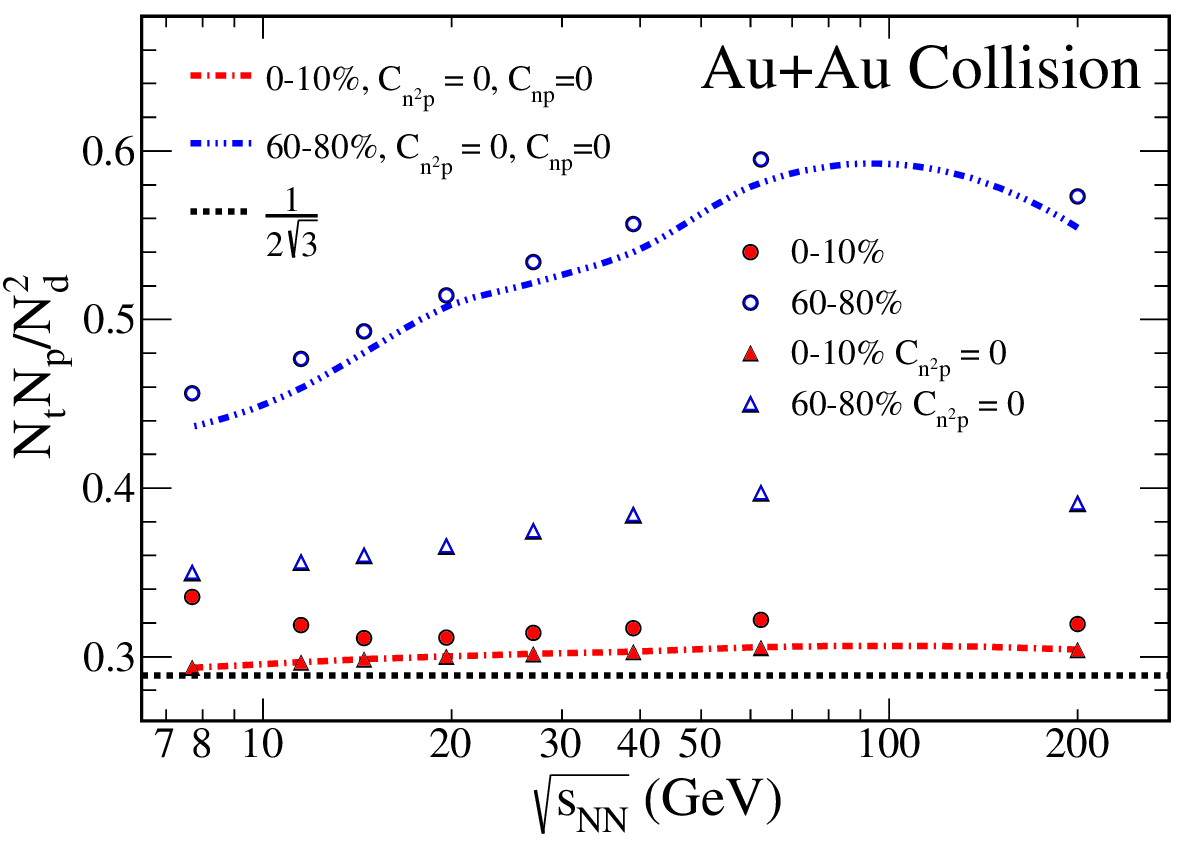}
    \caption{\label{fig4} Collision energy dependence of the light nuclei yield ratio $N_tN_p/N_d^2$ from AMPT for Au+Au collisions with $|y|\leqslant 0.5$. The results from 0\%-10\% central Au+Au collision are shown as solid circles by Eq.~(\ref{lnr}), solid triangle by Eq.~(\ref{lnr1}) and red dashed line by Eq.~(\ref{lnr2}). The results from 60\%-80\% peripheral Au+Au collision are shown as opened circles by Eq.~(\ref{lnr}), opened triangle by Eq.~(\ref{lnr1}) and blue dash-dot line by Eq.~(\ref{lnr2}).}
\end{figure}

Figure~\ref{fig4} illustrates the collision energy dependence of the light nuclei yield ratio $N_tN_p/N_d^2$ extracted from 0\%-10\% central and 60\%-80\% peripheral Au+Au collisions within $|y|\leqslant 0.5$. The dash-dot lines represent cases where $C_{np}$ and $C_{n^2p}$ are disregarded. A slightly increase in the light nuclei yield ratio with increasing collision energy is evident from the AMPT model. At 0\%-10\% central collisions, the yield ratios are consistent with predictions from the coalescence model calculations, $1/2\sqrt{3}$. Peripheral collisions show larger yield ratios in comparison to central collisions. When $C_{n^2p}$ is not considered, it leads to a reduction in the yield ratio for both central and peripheral collisions, which suggest that neglecting $C_{n^2p}$ would yield an overestimate of neutron density fluctuation from experimental data. Interestingly, the discrepancy between experimental signals and the true physical signals, induced by the omission of $C_{n^2p}$, remains unaffected by the collision energy, except at the lower energies, specifically at 7.7 GeV. At lower energy regime, the influence of $C_{n^2p}$ on the yield ratio becomes more important, whose underlying mechanisms are unclear and constitute a focal point for future research.

\begin{figure}[!htp]
    \includegraphics[width=0.8\textwidth]{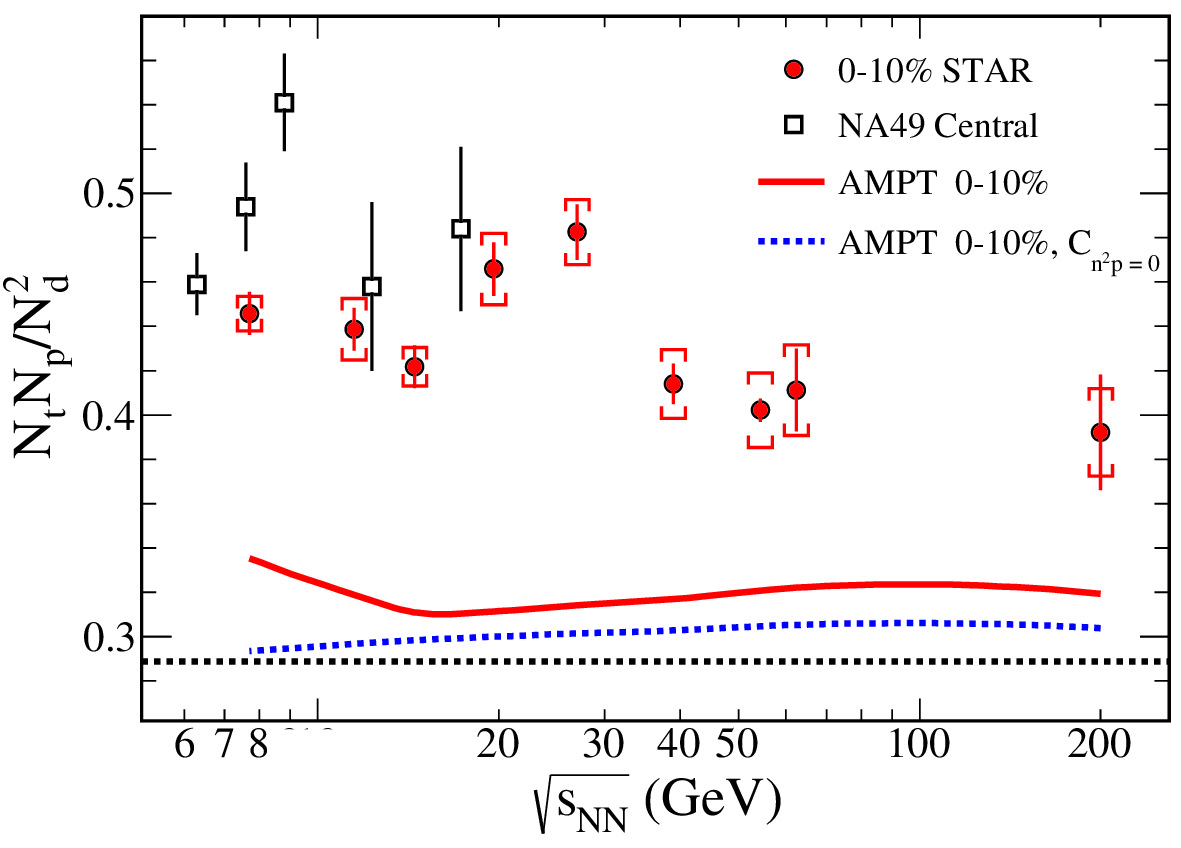}
    \caption{\label{fig5} Collision energy and centrality dependence of the light nuclei yield ratio $N_tN_p/N_d^2$ from AMPT with $|y|\leqslant 0.5$. Solid circles are the results from STAR detector at 0\%-10\% central Au+Au collision~\cite{RN218}. Open squares are the results from NA49 at central Pb+Pb collision~\cite{RN19,RN9}.}
\end{figure}

Fig.~\ref{fig5} compare the experimental results from STAR at 0\%-10\% central Au+Au collisions~\cite{RN218} and NA49 at central Pb+Pb collisions~\cite{RN19,RN9}. Since we compare the central collisions, the impact of $C_{np}$ on the yield from AMPT is deemed negligible. The inclusion of $C_{n^2p}$ enhances the light nuclei yield ratios, bringing AMPT model estimates closer to the experimental results. However, the inclusion of $C_{n^2p}$ does not change the collision energy dependence of the yield ratio, thereby failing to reproduce the non-monotonic behavior observed in experiment. Given the absence of critical phenomena in the AMPT, this result is reasonable. Through the AMPT model, we obtain a better baseline of the light nuclei yield ratio.

In summary, using the AMPT model for Au+Au collisions, we study the rapidity, collision energy, and centrality dependence of the relative neutron density fluctuation $\sigma_n/\langle n \rangle$, the two nucleons and three nucleons correlations $C_{np}$ and $C_{n^2p}$. The related light nuclei yield ratios from Eq.~(\ref{lnr}), Eq.~(\ref{lnr1}), and Eq.~(\ref{lnr2}) are also investigated. At central or mid-central collisions, the influence of $C_{np}$ on the light nuclei yield ratios is insignificant. However, in peripheral collisions, a non-zero $C_{np}$ will lead to a reduction in the light nuclei yield ratio. Importantly, regardless of whether in central or peripheral collisions, the $C_{n^2p}$ leads to an overall enhancement in the light nuclei yield ratios. Due to the absence of critical physics in the AMPT model, it fails to reproduce the experimental observations, particularly the peak observed in the light nuclei yield ratio around $\snn = $20-30 GeV. Incorporating the three-nucleon correlation $C_{n^2p}$, our model leads to results that offer a more accurate baseline, closer to the true experimental values.
%\section{}

%\begin{acknowledgments}
The authors appreciate the referee for his/her careful reading of the paper and valuable comments. This work is supported in part by the Scientific Research Foundation of Hubei University of Education for Talent Introduction (No. ESRC20230002 and No. ESRC20230007) and Research Project of Hubei Provincial Department of Education (No. D20233003 and No. B2023191).
%\end{acknowledgments}

\bibliography{apssamp}

\end{document}